\documentclass[letterpaper, 10 pt, conference]{ieeeconf}
\usepackage{graphics,epsfig,epstopdf,mathptmx,times,amsmath,amssymb,psfrag,mathtools}
\usepackage{cite}
\usepackage[utf8]{inputenc}
\usepackage[english]{babel}
\usepackage{mleftright}
\usepackage[makeroom]{cancel}
\usepackage{tikz}

\newcommand*\circled[1]{\tikz[baseline=(char.base)]{
            \node[shape=circle,draw,inner sep=1pt] (char) {#1};}}

\newtheorem{theorem}{Theorem}
\newtheorem{corollary}{Corollary}[theorem]
\newtheorem{lemma}[theorem]{Lemma}
\newtheorem{prop}{Proposition}
\newtheorem{remark}{Remark}

\title{\LARGE \bf Controllability and Accessibility Results for $N$-Link Horizontal Planar Manipulators with One Unactuated Joint}

\author{Tan Chen and Bill Goodwine$^{1}$% <-this % stops a space 
  \thanks{*The partial support of the US National Science Foundation
    under grant IIS-1527393 is gratefully
    acknowledged.}% <-this % stops a space  
\thanks{$^{1}$All the authors are with the Department of Aerospace \&
  Mechanical Engineering, University of Notre Dame, Notre Dame, IN
  46556 USA, {\tt\small tchen8@nd.edu, bill@controls.ame.nd.edu}.}} 

\begin{document}
\maketitle
\thispagestyle{empty}
\pagestyle{empty}

\begin{abstract}

This paper presents the accessibility and small-time local controllability (STLC) results for $N$-link horizontal planar manipulators with only one unactuated joint. STLC is important in controls, both for design considerations and because large and swinging maneuvers may be avoided for close reconfiguration if a system is STLC. Despite the fact that controllability of underactuated horizontal planar manipulators has been extensively studied, most work focused only on three-link and global controllability. This paper thus has two contributions: 1) using Lie brackets to study the accessibility and STLC for underactuated two-link manipulators with different actuator configurations, and illustrating the results from a perspective of system dynamics, 2) obtaining the accessibility and STLC results for $N$-link manipulators with one unactuated joint by considering realistic models and different actuator configurations. It is found that an $N$-link ($N\geq 3$) with the first joint actuated is STLC for a subset of equilibrium points based on Sussmann's general theorem for STLC. Moreover, with the dynamics of $N$-link considered in the controllability analysis, it gives relatively simple forms for the nontrivial vector fields, which make it easy to determine at which configurations a model loses full rank condition for accessibility. 

\end{abstract}

\section{Introduction}
\label{introduction}
Underactuated mechanical systems (UMS) \cite{liu2013survey} have been widely studied for several decades. Examples of UMS include spacecraft, snake robots, biped robots, underactuated manipulators etc. One of the most important challenges for UMS is the control problem, since many traditional control methods are not directly applicable in this area. Prior to controller design for an UMS, controllability should be investigated.

Control of a vertical two-link underactuated robots has been studied in \cite{block1995mechanical,spong1994partial,spong1995pendubot, spong1995swing} by presenting a partial feedback linearization method. The idea is to first apply the partial feedback linearization to drive the system to a basin of attraction of a linear regulator, and then switch to Linear Quadratic Regulator to balance the robot. The work in \cite{spong1996energy, fantoni2000energy} further considered the passivity properties of the systems and proposed energy-based control method that does not require high gains. It can be noted that these methods require the linearization-based controllability at the equilibrium. However, it is easily verified that the linearization of an underactuated horizontal manipulator is not controllable at the equilibrium because of the absense of a gravity force, which makes the control problem for a horizontal pendubot more difficult in some sense. 

Later research work thus turned more towards the nonlinear control theories to tackle the control problem for a horizontal underactuated manipulator. The work in \cite{de1997control} studied controllability for underactuated systems based on the control-affine form of the system and neglecting the model dynamics, and it provided sufficient conditions for STLC at the equilibrium. The work in\cite{kobayashi2002controllability} exploited the time reversal property and showed that a horizontal manipulator with one unactuated joint is controllable, \textit{i.e.}, there exists an admissible trajectory to drive a state to any other state, if and only if the first joint is actuated. The work in \cite{arai1998nonholonomic} showed STLC for a three-link horizontal pendubot by assuming a simplified model and neglecting the dynamics of the first and second links, which thus converting the original problem to a STLC problem for an underactuated hovercraft. In contrast, the work in \cite{mahindrakar2005controllability} also focused on three-link horizontal planar underactuated manipulators but discussed the STLC for models at various actuator configurations with the dynamics considered. The work in \cite{lynch2000collision,bullo2001kinematic} introduced a new notion of kinematic controllability for underactuated mechanical systems and proposed the decoupling vector fields method, which was implemented on planning a collision-free trajectory for a three-link horizontal pendubot. 

Note that most of the previous studies focused on three-link and global controllability. Even though STLC was discussed in some literatures, the controllability analysis was based on a control-affine form, with model dynamics neglected. The work in \cite{chen2019controllability} thus extended the previous controllability results to a more general case, by considering a realistic $N$-link horizontal pendubot and exploiting Lie brackets to study the accessibility and STLC for the model. This paper considers both pendubot and acrobot, \textit{i.e.}, a horizontal planar manipulator with one unactuated joint that may occur at any position. Through comparison and analysis, it provides more complete and insightful results about the controllability for underactuated horizontal manipulators. Section \ref{sec:nonlinear} provides a brief review on the tools for studying nonlinear controllability. Section \ref{sec:twolink} and \ref{sec:threelink} demonstrate the main results of the paper.
\section{Introduction to Nonlinear Controllability}
\label{sec:nonlinear}
Consider a control-affine nonlinear system~\cite{nijmeijer1990nonlinear}
\begin{equation}
\label{eq:affinesys}
\dot{x} = f(x)+\sum_{i=1}^{m}g_i(x)u_i,
\end{equation}
where $x=(x_1,\dotsc,x_n)$ are local coordinates for a smooth manifold $\mathcal{M}$ (the state space manifold), $u=(u_1,\dotsc,u_m)$ are input controls, and $f,g_1,\dotsc,g_m$ are smooth vector fields on $\mathcal{M}$. The principle question about controllability of a nonlinear system is whether there exists an admissible trajectory from any given initial point to any final point on $\mathcal{M}$ with a suitable choice of control inputs. The simplest way to study it is to consider linearization of the system at equilibrium, since the controllability for a linear system can be easily verified with Kalman rank condition~\cite{antsaklis2007linear}.

However, the linearization lacks much of the structure of the original system, and it occurs often that a nonlinear system is controllable while its lineariation is not.  The \emph{accessibility algebra} $\mathcal{C}$ for the system~(\ref{eq:affinesys}) is the smallest algebra of the Lie algebra of vector fields on $\mathcal{M}$ that contains $f,g_1,\dotsc,g_m$. Define the \emph{accessibility distribution}, $\varDelta$, generated from the accessibility algebra $\mathcal{C}$,
\begin{equation*}
\varDelta(x) = \operatorname{span}\{X(x)|X\in \mathcal{C}\},
\end{equation*}
where $X$ is a vector field on $\mathcal{M}$. A system is locally accesible from $x_0 \in \mathbb{R}^n$ if the set of reachable states from $x_0$ within time $T$ has a non-empty interior for all $T>0$. 

\begin{theorem}[from \cite{nijmeijer1990nonlinear}]
\label{thm:distribution}
The system~(\ref{eq:affinesys}) is locally accessible from $x_0$ if and only if
\begin{equation*}
    \operatorname{dim}\:\varDelta(x_0) = n.
\end{equation*}
\end{theorem}
Moreover, the distribution $\varDelta$ is said to be \emph{regular} \cite{murray2017mathematical} if the dimension of the subspace $\varDelta(x)$ does not vary with $x$. For a system that is locally accessible from almost any state, it may lose the full rank condition for accessibility at some states, which we name \emph{singular} states.

Accessibility is a necessary but not sufficient condition for controllability. Sussmann~\cite{sussmann1987general} provided sufficient conditions for small-time local controllability (STLC) of a system. The definition of STLC from $x_0$ is that the set of reachable states from $x_0$  within time $T$ has a non-empty interior for all $T>0$ and the state $x_0$ is in the interior. Furthermore, if a system is STLC from all states in the manifold $\mathcal{M}$, the system can be controlled on $\mathcal{M}$ by designing a STLC-based controller to connect from any intial state to any final state. 

Consider a Lie bracket $B$ generated from the vector fields on the manifold $\mathcal M$, and $\delta^0(B),\delta^1(B),\dotsc,\delta^m(B)$ represent the occurrence numbers of the vector fields $f,g_1,\dotsc,g_m$ in $B$ respectively. The bracket $B$ is bad if $\delta^0(B)$ is odd and $\delta^1(B),\dotsc,\delta^m(B)$ are all even (including zero). A bracket is good if it is not bad. A $\theta$-degree $\delta_{\theta}(B)$ is defined for the vector field $B$ by~\cite{bianchini1993controllability}
\begin{equation*}
\delta_{\theta}(B) = \sum_{j=0}^{m}\theta_j\delta^j(B),
\end{equation*}
where the numbers $\theta_0,\theta_1,\dotsc,\theta_m$ satisfy $\theta_j \geq \theta_0 \geq 0$, $j=1,\dotsc,m$. A bad bracket is said to be $\theta$-neutralized if it is a linear combination of lower $\theta$-degree good brackets. 

\begin{theorem}[from \cite{bianchini1993controllability}]
\label{thm:stlc}
A system is STLC from an equilibrium point if there exist a sufficient number of good brackets at the equilibrium point to span the full-dimensional space and all bad brackets are $\theta$-neutralized.
\end{theorem}

The following theorem provides a criterion for the nontrivial brackets at the equilibrium states. The proof can be found in~\ref{sec:appendix-B}, which follows the same line with \cite{lewis1997configuration}.
\begin{theorem}
\label{nontrivialbrackets}
For the $N$-link horizontal planar manipulators with one unactuated joint, the only nontrivial brackets evaluated at the equilibrium are those satisfying
\begin{equation*}
\sum_{i=1}^m \delta^i(B)-\delta^0(B) = 0\, \operatorname{or} \, 1.
\end{equation*}
\end{theorem}

\section{Two-Link Model}
\label{sec:twolink}
\begin{figure}
    \centering
    \includegraphics[width = .4\textwidth]{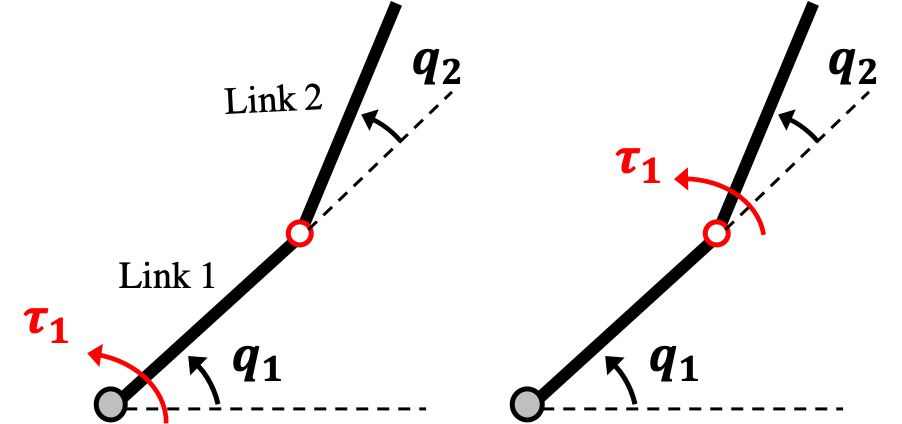}
    \caption{Two-link horizontal manipulators. No gravity concerned in the dynamics. Left: pendubot, right: acrobat. The masses, moments of inertia, link lengths and distances between center of masses and corresponding joints are $m_1$, $I_1$, $l_1$, $l_{c_1}$ for the link 1, and $m_2$, $I_2$, $l_2$, $l_{c_2}$ for the link 2.}
    \label{figs:twolink}
\end{figure}

To set up the general $N$-link results, this section presents accessibility and STLC results for two-link horizontal manipulators with one degree of unactuation. The two-link has two possible configurations as illustrated in Fig. \ref{figs:twolink}. Define:
\begin{equation*}
    \alpha_1 = m_1l_{c_1}^2+m_2l_1^2+I_1, \quad
    \alpha_2 = m_2l_{c_2}^2+I_2,
    \quad
    \beta_1 = m_2l_1l_{c_2}.
\end{equation*}

\subsection{Pendubot Configuration}
The dynamics for a horizontal pendubot are described by
\begin{equation}
\label{eq:pendubot}
    \begin{aligned}
    \begin{bmatrix}
    M_{11} & M_{12} \\
    M_{21} & M_{22}
    \end{bmatrix}
    \begin{bmatrix}
    \ddot{q}_1 \\
    \ddot{q}_2
    \end{bmatrix}
    +
    \begin{bmatrix}
    C_1 \\ C_2
    \end{bmatrix}
    =
    \begin{bmatrix}
    \tau_1 \\ 0
    \end{bmatrix}
    \end{aligned},
\end{equation}
where in the inertia matrix, \footnote{$s_2$ and $c_2$ are abbreviations for $\operatorname{sin}(q_2)$ and $\operatorname{cos}(q_2)$, respectively.}
\begin{equation*}
\begin{aligned}
    M_{11} &= \alpha_1+\alpha_2+2\beta_1c_2 \quad
    &M_{12} &= \alpha_2+\beta_1c_2 \\
    M_{21} &= \alpha_2+\beta_1c_2 
    &M_{22} &= \alpha_2
\end{aligned}
\end{equation*}
and in the Coriolis matrix,
\begin{equation*}
    C_1 = -2\beta_1\dot{q}_1\dot{q}_2s_2-\beta_1\dot{q}_2^2s_2 \quad
    C_2 = \beta_1\dot{q}_1^2s_2.
\end{equation*}

In order to simplify the nonlinear controllability analysis, partial feedback linearization~\cite{spong1994partial} is performed to transform the model dynamics to a simpler form. We introduce an input $u_1$. The control input $\tau_1$ can thus be designed by
\begin{equation}
    \tau_1 = \frac{M_{22}C_1-M_{12}C_2}{M_{22}}+\frac{M_{11}M_{22}-M_{12}M_{21}}{M_{22}}u_1,
\end{equation}
which yields to 
\begin{equation}
\begin{aligned}
    \ddot{q}_1 &= u_1 \\
    \ddot{q}_2 &= -\frac{C_2}{M_{22}}-\frac{M_{21}}{M_{22}}u_1.
\end{aligned}
\end{equation}

Let $x_1 = q_1$, $x_2 = q_2$, $x_3 = \dot{q}_1$ and $x_4 = \dot{q}_2$, which gives,
\begin{equation}
    \dot{x} = f(x) + g_1(x)u_1,
\end{equation}
where the state vector $x = 
(x_1,\,x_2,\,x_3,\,x_4)$, the drift field $f(x) = 
(x_3,\,x_4,\,0,\,-C_2/M_{22})$, and the input vector field $g_1(x) = 
(0,\,0,\,1,\,-M_{21}/M_{22})$. The equilibrium points are zero-velocity states, where $\dot{q}_1 = \dot{q}_2 = 0$.

\begin{theorem}
\label{thm:paccess}
A two-link horizontal pendubot is accessible from almost any state. 
\end{theorem}

\begin{proof}
Consider
\begin{equation}
\label{eq:pendubotvecfields}
\begin{aligned}
g_1(x) &= 
(0,\,0,\,1,\,A_1(x))\\
[f,g_1](x)
&=
(-1,\,-A_1(x),\,0,\,A_2(x))\\
[g_1,[f,g_1]](x)
&=
(0,\,0,\,0,\,A_3(x))\\
[f,[g_1,[f,g_1]]](x)
&=
(0,\,-A_3(x),\,0,\,A_4(x)),
\end{aligned}
\end{equation}
where
\begin{equation*}
\begin{aligned}
    A_1(x) &= -\frac{\alpha_2+\beta_1\operatorname{cos}(x_2)}{\alpha_2} \\
    A_2(x) &= \frac{\beta_1(2x_3+x_4)\operatorname{sin}(x_2)}{\alpha_2}\\
    A_3(x) &= 
    -\frac{\beta_1^2\operatorname{sin}(2x_2)}{\alpha_2^2}\\
    A_4(x) &=
    -\frac{2\beta_1^2x_4\operatorname{cos}(2x_2)}{\alpha_2^2}.
    \end{aligned}
\end{equation*}

When $A_3(x) \neq 0$, \textit{i.e.}, $x_2 \neq k\pi/2$, $k \in \mathbb{Z}$, the vector fields
\begin{equation*}
\underbrace{
\mleft[
\begin{array}{c}
0 \\
0 \\
\hline
1 \\
A_1(x)
\end{array}
\mright]
}_{g_1}
\quad
\underbrace{
\mleft[
\begin{array}{c}
0 \\
0 \\
\hline
0 \\
A_3(x)
\end{array}
\mright]
}_{[g_1,[f,g_1]]}
\quad
\underbrace{
\mleft[
\begin{array}{c}
-1 \\
-A_1(x) \\
\hline
0 \\
A_2(x)
\end{array}
\mright]
}_{[f,g_1]}
\quad
\underbrace{
\mleft[
\begin{array}{c}
0 \\
-A_3(x) \\
\hline
0 \\
A_4(x)
\end{array}
\mright]
}_{[f,[g_1,[f,g_1]]]}
\end{equation*}
are independent and span the state space.
\end{proof}

Accessibility is necessary condition for controllability, and~\cite{kobayashi2002controllability} has proven that such horizontal pendubot is completely controllable, \textit{i.e.}, there exists an admissible trajectory from any given initial point to any given final point. However, we will show that the pendubot does not satisfy the sufficient conditions for STLC in these coordinates. Recall that STLC is not a coordinate invariant property \cite{bullo2004geometric}.

\begin{theorem}
The two-link horizontal pendubot does not satisfy the sufficient conditions for STLC stated in Theorem~\ref{thm:stlc}.
\end{theorem}

\begin{remark}
The nontrivial brackets at equilibrium are,
\begin{equation*}
\begin{aligned}
    \text{Degree 1:}& \quad \{f,\,g_1\}\\
    \text{Degree 2:}& \quad \{[f,g_1]\}\\
    \text{Degree 3:}& \quad \{[g_1,[f,g_1]\}\\
    \text{Degree 4:}& \quad \{[f,[g_1,[f,g_1]]],\,[g_1,[f,[f,g_1]]]\,\}.\\
\end{aligned}
\end{equation*}
When the degree is over $3$, the nontrivial brackets at equilibrium should contain at least two $f$ and two $g_1$. Otherwise, the brackets are trivial at equilibrium by Theorem~\ref{nontrivialbrackets}, such as $[f,f,[f,g_1]]$ and $[g_1,[g_1,[f,g_1]]]$ are trivial at equilibrium.
\end{remark}

\begin{proof}
Of the four vector fields in Eq.~(\ref{eq:pendubotvecfields}), $[g_1,[f,g_1]$ is a bad bracket. The goal is to find good brackets with lower $\theta$-degree to neutralize $[g_1,[f,g_1]]$. Obviously it cannot be $\theta$-neutralized by $g_1$ and $[f,g_1]$, since they are independent except when $A_3(x) = 0$ making $[g_1,[f,g_1]]$ trivial. 

Using Theorem~\ref{nontrivialbrackets}, the nontrivial good brackets (evaluated at the equilibrium) except $g_1$, $[f,g_1]$ and $[f,[g_1,[f,g_1]]]$ should contain at least two $g_1$ and two $f$, leading to a $\theta$-degree no smaller than that for $[g_1,[f,g_1]]$. Therefore $[g_1,[f,g_1]]$ cannot be $\theta$-neutralized, and thus the two-link horizontal pendubot does not satisfy the sufficient conditions for STLC.
\end{proof}

\subsection{Acrobat Configuration}
The dynamics for a horizontal acrobat are described by
\begin{equation}
\label{eq:modelacrobat}
    \begin{aligned}
    \begin{bmatrix}
    M_{11} & M_{12} \\
    M_{21} & M_{22}
    \end{bmatrix}
    \begin{bmatrix}
    \ddot{q}_1 \\
    \ddot{q}_2
    \end{bmatrix}
    +
    \begin{bmatrix}
    C_1 \\ C_2
    \end{bmatrix}
    =
    \begin{bmatrix}
    0 \\ \tau_1
    \end{bmatrix}
    \end{aligned},
\end{equation}
where the inertia and Coriolis matrices are the same with those in Eq.~(\ref{eq:pendubot}).

We introduce an input $u_1$, and design the control input $\tau_1$ by
\begin{equation}
    \tau_1 = \frac{M_{11}C_2-M_{21}C_1}{M_{11}}+\frac{M_{11}M_{22}-M_{12}M_{21}}{M_{11}}u_1,
\end{equation}
which yields to
\begin{equation}
    \begin{aligned}
        \ddot{q}_1 &= -\frac{C_1}{M_{11}}-\frac{M_{12}}{M_{11}}u_1 \\
        \ddot{q}_2 &= u_1.
    \end{aligned}
\end{equation}

Let $x_1 = q_1$, $x_2 = q_2$, $x_3 = \dot{q}_1$ and $x_4 = \dot{q}_2$, which gives, 
\begin{equation}
    \dot{x} = f(x)+g_1(x)u_1,
\end{equation}
where the state vector
$x = 
(x_1,\,x_2,\,x_3,\,x_4)
\triangleq
(q_1,\,q_2,\,\dot{q}_1,\,\dot{q}_2)
$, the drift field $f(x) =
(x_3,\,x_4,\,-C_1/M_{11},\,0)$,
and the input vector field $g_1(x) = 
(0,\,0,\,-M_{12}/M_{11},\,1)$.

\begin{lemma}
\label{lm:involutive}
Let $\varDelta$ be be a non-singular distribution on $\mathcal{M}$ generated by vector fields $f_1, \dotsc ,f_r$. $\varDelta$ is involutive if and
only if $[f_i,f_j] \in \varDelta$ for all $1 \leq i, j \leq r$. 
\end{lemma}

\begin{proof}
1) Necessity can be easily proven by the definition of ``involutive distribution''.

2) Sufficiency: Pick any two vector fields $k_1, k_2 \in \varDelta$,
\begin{equation*}
\begin{aligned}
k_1 &= \sum_{p=1}^r \alpha_p f_p \\
k_2 &= \sum_{q=1}^r \beta_q f_q,
\end{aligned}
\end{equation*}
which yield to the bracket between these two vector fields,
\begin{equation*}
[k_1,k_2] =  \sum_{p=1}^r\sum_{q=1}^r\alpha_p  \beta_q[f_p,f_q].
\end{equation*}

Since $[f_i,f_j] \in \varDelta$ for all $1 \leq j, j\leq r$, we can conclude $[k_1,k_2] \in \varDelta$, and thus the distribution $\varDelta$ is involutive. 
\end{proof}

\begin{theorem}
A two-link horizontal acrobat is not accessible from any state.
\end{theorem}

\begin{proof}
The first-order Lie brackets for the model are
\begin{equation*}
\begin{aligned}
f(x) &= 
(x_3,\,x_4,\,A_1(x),\,0)\\
g_1(x) &= 
(0,\,0,\,A_2(x),\,1),
\end{aligned}
\end{equation*}
the \emph{nontrivial} second-order brackets are
\begin{equation*}
[f,g_1](x) = 
(-A_2(x),\,-1,\,-A_1(x)/x_4,\,0),
\end{equation*}
and the \emph{nontrivial} third-order brackets are
\begin{equation*}
[f,[f,g_1]](x) = 
(A_3(x),\,0,\,0,\,0),
\end{equation*}
where
\begin{equation*}
\begin{aligned}
A_1(x) &= \frac{2 \beta_1 x_3 x_4 \operatorname{sin}(x_2) + \beta_1 x_4 ^2 \operatorname{sin}(x_2)}{\alpha_1 + \alpha_2 + 2 \beta_1 \operatorname{cos}(x_2)} \\
A_2(x) &= -\frac{\alpha_2+\beta_1\operatorname{cos}(x_2)}{\alpha_1+\alpha_2+2\beta_1\operatorname{cos}(x_2)}\\ 
A_3(x) &= \frac{2\beta_1 (\alpha_1 x_3+\alpha_2 (x_3+x_4) +\beta_1 (2x_3 + x_4) \operatorname{cos}(x_2))\operatorname{sin}(x_2)}{(\alpha_1+\alpha_2+2\beta_1\operatorname{cos}(x_2))^2}.
\end{aligned} 
\end{equation*}

Note that the bracket $[f,[f,g_1]]$ is a linear combination of two lower-order vector fields $f$ and $[f,g_1]$ by
\begin{equation*}
[f,[f,g_1]] = \frac{A_3}{-x_4A_2+x_3}\left(f + x_4 \cdot [f,g_1]\right).
\end{equation*}

Using Lemma~\ref{lm:involutive}, the distribution $\varDelta$ generated by $f$, $g_1$ and $[f,g_1]$ is involutive. The dimension of the distribution
\begin{equation*}
    \operatorname{dim}(\varDelta) = 3,
\end{equation*}
which is smaller than the dimension of the state space manifold $\mathcal{M}$. Therefore, the two-link horizontal acrobat is not accessible from any state.
\end{proof}

In fact, if we consider the distribution only on the manifold of all equilibrium points ($x_3 = x_4 = 0$), the dimension of the distribution is two. The results are consistent with the results in~\cite{oriolo1991control}. For a horizontal acrobat, because $q_1$ is a cyclic coordinate (which does not enter the inertia matrix) and no gravity is concerned in the dynamics, the first row in Eq.~(\ref{eq:modelacrobat}),
\begin{equation}
\label{eq:2ndnonholonomic}
    M_{11}\ddot{q}_1+M_{12}\ddot{q}_2+C_1=0
\end{equation}
can be partially integrated~\cite{oriolo1991control} to
\begin{equation}
\label{eq:1stnonholonomic}
    M_{11}\dot{q}_1+M_{12}\dot{q}_2+k_1=0,
\end{equation}
where $k_1$ is constant.

Eq.~(\ref{eq:1stnonholonomic}) provides a constraint on the velocity states, which can also be derived by momentum conservation. This explains why the dimension of the distribution $\varDelta$ is 3 in general. Further considering only the equilibrium points, $k_1=0$ in this case. Eq.~(\ref{eq:1stnonholonomic}) can be further integrated to a holonomic constraint,
\begin{equation}
\label{eq:holonomic}
q_2 = \frac{\alpha_1-\alpha_2}{2}\operatorname{arctan}(\frac{\sqrt{(\alpha_1+\alpha_2)^2-4\alpha^2_3}}{\alpha_1+\alpha_2+2\alpha_3}\operatorname{tan}\frac{q_2}{2})+k_2,
\end{equation}
where $k_2$ is constant. Hence, when evaluating on the manifold of equilibrium points, the dimension of $\varDelta$ is reduced to 2.

\begin{corollary}
The two-link horizontal acrobat is not STLC from any state.
\end{corollary}

\section{Three-Link to N-Link model}
\label{sec:threelink}
\subsection{Three-link Analysis}
\begin{figure}
    \centering
    \includegraphics[width = .5\textwidth]{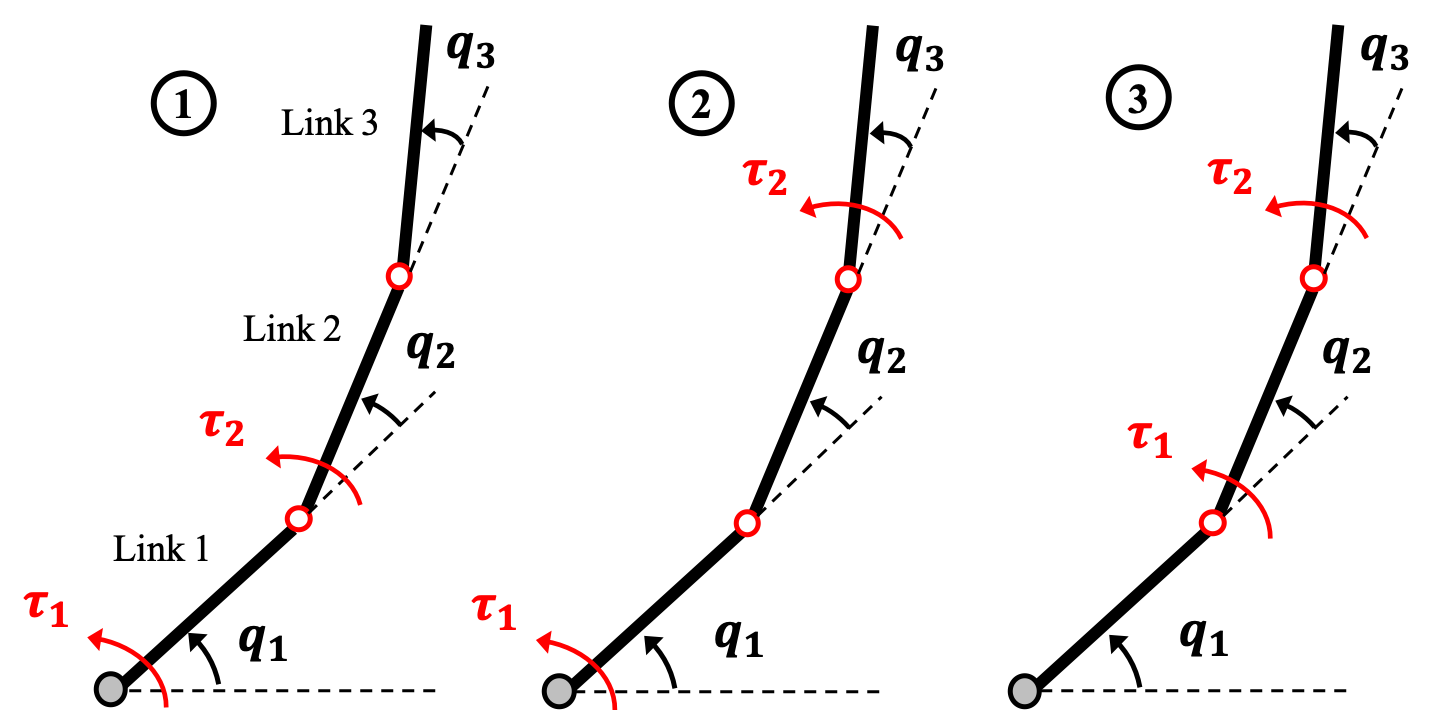}
    \caption{Three-link horizontal manipulators. No gravity concerned in the dynamics. Configurations 1 and 2: actuated at least at the first joint, configuration 3: unactuated at the first joint. The masses, moments of inertia, link lengths and distances between center of masses and corresponding joints are $m_1$, $I_1$, $l_1$, $l_{c_1}$ for the link 1, $m_2$, $I_2$, $l_2$, $l_{c_2}$ for the link 2, and $m_3$, $I_3$, $l_3$, $l_{c_3}$ for the link 3.}
    \label{fig:threelink}
\end{figure}

The configurations of three-link horizontal manipulators are shown in Figure~\ref{fig:threelink}. The work in \cite{kobayashi2002controllability} has proven that a horizontal three-link with the first joint actuated is controllable, while a three-link with the first joint unactuated is not. We will briefly present the accessibility and STLC results for the three configurations. Modeling for the three-link, underactuated horizontal manipulators and applying partial feedback linearization to simplify the equations can be found in~\ref{sec:appendix-A}. The notions for some moments of inertia can also be found in~\ref{sec:appendix-A}.

\subsubsection{Configuration $1$}
The dynamics can be written in control-affine form
\begin{equation}
\label{eq:config-1}
\dot{x} =  f(x)+g_1(x)u_1+g_2(x)u_2,
\end{equation}
where 
$x = 
(x_1,\,x_2,\,x_3,\,x_4,\,x_5,\,x_6)
\triangleq
(q_1,\,q_2,\,q_3,\,\dot{q}_1,\,\dot{q}_2,\,\dot{q}_3)$, $f(x) =(x_4,\,x_5,\,x_6,\,0,\,0,\,P(x))$,
$g_1(x) = (0,\,0,\,0,\,1,\,0,\,Q_1(x))$, and $g_2(x) = (0,\,0,\,0,\,0,\,1,\,Q_2(x))$, and
\begin{equation*}
\begin{aligned}
P(x) &= -\frac{x_4^2(\beta_3\operatorname{sin}(x_3)+\beta_4\operatorname{sin}(x_2+x_3))+2x_4x_5\beta_3\operatorname{sin}(x_3)}{\alpha_3} \\
& \quad -\frac{\beta_3x_5^2\operatorname{sin}(x_3)}{\alpha_3} \\
Q_1(x) &= -\frac{\alpha_3+\beta_3\operatorname{cos}(x_3)+\beta_4\operatorname{cos}(x_2+x_3)}{\alpha_3} \\
Q_2(x) &= -\frac{\alpha_3+\beta_3\operatorname{cos}(x_3)}{\alpha_3}.
\end{aligned}
\end{equation*}

\begin{theorem}
\label{thm:3linkaccessible}
The three-link horizontal pendubot with configuration 1 is accessible from almost any state. 
\end{theorem}

\begin{proof}
We can pick six vector fields as follows,
\begin{equation}
\label{eq:6brackets}
\begin{aligned}
g_1(x) &= (0,\,0,\,0,\,1,\,0,\,Q_1(x)) \\ 
g_2(x) &= (0,\,0,\,0,\,0,\,1,\,Q_2(x)) \\
[f,g_1](x) &= (-1,\,0,\,-Q_1(x),\,0,\,0,\,*) \\
[f,g_2](x) &= (0,\,-1,\,-Q_2(x),\,0,\,0,\,*) \\
[g_1,[f,g_2]](x) &= (0,\,0,\,0,\,0,\,0,\,R_1(x)) \\
[f,[g_1,[f,g_2]]](x) &= (0,\,0,\,-R_1(x),\,0,\,0,\,*)
\end{aligned}
\end{equation}
where
\begin{equation*}
R_1(x) = -\frac{\beta_3(\beta_3\operatorname{sin}(2x_3)+\beta_4\operatorname{sin}(x_2+2x_3))}{\alpha_3^2},
\end{equation*}
and $*$ represents an arbitrary expression.

When $R_1(x) \neq 0$, \textit{i.e.}, $\beta_3\operatorname{sin}(2x_3)+\beta_4\operatorname{sin}(x_2+2x_3) \neq 0$, the above six vector fields
\begin{equation*}
\begin{aligned}
&
\underbrace{
\mleft[
\begin{array}{c}
0 \\
0 \\
0 \\
\hline
1 \\
0 \\
Q_1(x)
\end{array}
\mright]
}_{g_1}
\quad
\underbrace{
\mleft[
\begin{array}{c}
0 \\
0 \\
0 \\
\hline
0 \\
1 \\
Q_2(x)
\end{array}
\mright]
}_{g_2}
\quad
\underbrace{
\mleft[
\begin{array}{c}
0 \\
0 \\
0 \\
\hline
0 \\
0 \\
R_1(x)
\end{array}
\mright]
}_{[g_1,[f,g_2]]}
\\
&
\underbrace{
\mleft[
\begin{array}{c}
-1 \\
0 \\
-Q_1(x) \\
\hline
0 \\
0 \\
*
\end{array}
\mright]
}_{[f,g_1]}
\quad
\underbrace{
\mleft[
\begin{array}{c}
0 \\
-1 \\
-Q_2(x) \\
\hline
0 \\
0 \\
*
\end{array}
\mright]
}_{[f,g_2]}
\quad
\underbrace{
\mleft[
\begin{array}{c}
0 \\
0 \\
-R_1(x) \\
\hline
0 \\
0 \\
*
\end{array}
\mright]
}_{[f,[g_1,[f,g_2]]]}
\end{aligned}
\end{equation*}
are independent by inspection and span the state space.
\end{proof}

\begin{remark}Some of the singular states include cases when $x_2 = k_1\pi$ and $x_3 = k_2\pi/2$ where $k_1,\,k_2 \in \mathbb{Z}$. 
\end{remark}

Accessibility only guarantees that the dimension of reachable space from the states is full dimension. We further check STLC that enables changing the states in all directions within any time $T>0$ for three-link horizontal pendubots. 

\begin{theorem}
\label{thm:stlc3link}
The three-link horizontal pendubot with configuration 1 as described by Eq. (\ref{eq:threelinkconfig1}) is STLC from a subset of equilibrium points. 
\end{theorem}

\begin{proof}
Note that all the brakcets in Eq.~(\ref{eq:6brackets}) are good. We have to show that all bad Lie brackets are $\theta$-neutralized. Two bad brackets that have the same number of vector fields as $[g_1,[f,g_2]]$ are
\begin{equation*}
\begin{aligned}
[g_1,[f,g_1]](x) &= (0,\,0,\,0,\,0,\,0,\,R_2(x)) \\
[g_2,[f,g_2]](x) &= (0,\,0,\,0,\,0,\,0,\,R_3(x)),
\end{aligned}
\end{equation*}
where
\begin{equation*}
\begin{aligned}
R_2(x) &= -\frac{\beta_4(\beta_4\operatorname{sin}(2(x_2+x_3))+2\beta_3\operatorname{sin}(x_2+2x_3))}{\alpha_3^2} \\
& \quad -\frac{\beta_3^2\operatorname{sin}(2x_3)}{\alpha_3^2} \\
R_3(x) &= -\frac{\beta_3^2\operatorname{sin}(2x_3)}{\alpha_3^2}. 
\end{aligned}
\end{equation*}

Since they cannot be neutralized by $[g_1,[f,g_2]]$ at the same time, we assign the $\theta$-degrees to $f$, $g_1$ and $g_2$ by setting $\theta_0 = 1$,  $\theta_1 = 1$ and $\theta_2 = 2$,
which will give the $\theta$-degrees $4$, $3$, $5$ for $[g_1,[f,g_2]]$, $[g_1,[f,g_1]]$ and $[g_2,[f,g_2]]$ respectively. This indicates that the bad bracket $[g_2,[f,g_2]]$ can be neutralized by the good bracket $[g_1,[f,g_2]]$, while $[g_1,[f,g_1]]$ cannot. Therefore, we have to require $[g_1,[f,g_1]]$ to be trivial to make the $\theta$-neutralization happen for the two bad brackets. Hence, the maximum $\theta$-degree for all vector fields in Eq.~(\ref{eq:6brackets}) is 5. 

Note that the two bad brackets $[f,[f,[g_1,[f,g_1]]]]$ and $[g_1,[g_1,[g_1,[f,g_1]]]]$ also have 5 as the $\theta$-degree, but by using Theorem \ref{nontrivialbrackets} they are equal to zero when evaluated at the equilibrium. The $\theta$-degree for all other bad brackets is greater than 5 and can thus be neutralized.

Another way to assign the $\theta$-degrees to $f$, $g_1$ and $g_2$ is
$\theta_0 = 1$, $\theta_1 = 2$, and $\theta_2 = 1$,
which will give the $\theta$-degrees $4,\,5,\,3$ for $[g_1,[f,g_2]]$, $[g_1,[f,g_1]]$ and $[g_2,[f,g_2]]$ respectively. In this case, we need to make $[g_2,[f,g_2]]$ trivial in order to neutralize both bad brackets. Thus the system is STLC from all equilibrium states $x_e$ satisfying $R_2(x_e)R_3(x_e) = 0$ and $R_1(x_e) \neq 0$. 
\end{proof}

\begin{remark}
The pendubot with configuration $1$ is STLC from the equilibrium states when $x_3 = k_1 \pi/2$ and $x_2 \neq k_2 \pi$, $k_1,\,k_2 \in \mathbb{Z}$. Theorem~\ref{thm:stlc3link} does not claim that the system is not STLC from the equilibrium points that fail to satisfy $R_2(x_e)R_3(x_e) = 0$ and $R_1(x_e) \neq 0$, since Sussmann's general theorem provides sufficient conditions on STLC, which is also a stronger property than controllability.
\end{remark}

\subsubsection{Configuration $2$}
Following the same analysis for configuration $1$, it can be shown that the three-link horizontal pendubot with configuration $2$ is accessible from almost any state and STLC from a subset of equilibrium points. 

\subsubsection{Configuration $3$}
The dynamics are described by
\begin{equation}
\dot{x} =  f(x)+g_1(x)u_1+g_2(x)u_2,
\end{equation}
where 
$x = 
(x_1,\,x_2,\,x_3,\,x_4,\,x_5,\,x_6)
\triangleq
(q_1,\,q_2,\,q_3,\,\dot{q}_1,\,\dot{q}_2,\,\dot{q}_3)$, $f(x) =(x_4,\,x_5,\,x_6,\,S(x),\,0,\,0)$,
$g_1(x) = (0,\,0,\,0,\,T_1(x),\,1,\,0)$, and $g_2(x) = (0,\,0,\,0,\,T_2(x),\,0,\,1)$, and
\begin{equation*}
\begin{aligned}
S(x) = \frac{\operatorname{Num}_1}{\operatorname{Den}}, \,T_1(x) =\frac{\operatorname{Num}_2}{\operatorname{Den}},\,
T_2(x) =\frac{\operatorname{Num}_3}{\operatorname{Den}},
\end{aligned}
\end{equation*}
and
\begin{equation*}
\begin{aligned}
\operatorname{Num}_1 &=
\beta_4 (x_5 + x_6)(2 x_4 + 2 x_5 + x_6)\operatorname{sin}(x_2 + x_3) \\
& \quad +\beta_3 x_6(2 x_4 + 2 x_5 + x_6)\operatorname{sin}(x_3)\\
& \quad +(\beta_1 + \beta_2) (2 x_4 + x_5) x_5 \operatorname{sin}(x_2) \\
\operatorname{Num}_2 &= -\alpha_2 - \alpha_3 - (\beta_1 + \beta_2) \operatorname{cos}(x_2) - 2\beta_3 \operatorname{cos}(x_3) \\
& \quad - \beta_4 \operatorname{cos}(x_2 + x_3) \\
\operatorname{Num}_3 &= -\alpha_3 - \beta_3 \operatorname{cos}(x_3) - \beta_4 \operatorname{cos}(x_2 + x_3) \\
\operatorname{Den} &= \alpha_1+\alpha_2+\alpha_3 +2(\beta_1+\beta_2)\operatorname{cos}(x_2)+2\beta_3\operatorname{cos}(x_3) \\
& \quad +2\beta_4\operatorname{cos}(x_2+x_3).
\end{aligned}
\end{equation*}

\begin{theorem}
\label{thm:3linkacrobataccessible}
The three-link horizontal manipulator with configuration 3 is not accessible from any state. 
\end{theorem}

\begin{proof}
Consider a distribution,
\begin{equation*}
\varDelta = \{f,\,g_1,\,g_2,\,[f,g_1],\,[f,g_2]\},
\end{equation*}
where 
\begin{equation*}
\begin{aligned}
[f,g_1](x) &= [-T_1(x),\,-1,\,0,\,V_1(x),\,0,\,0] \\
[f,g_2](x) &= [-T_2(x),\,0,\,-1,\,V_2(x),\,0,\,0],
\end{aligned}
\end{equation*}
and
\begin{equation*}
V_1(x) = \frac{\operatorname{Num}_4}{\operatorname{Den}},\,V_2(x) = \frac{\operatorname{Num}_5}{\operatorname{Den}}
\end{equation*}
\begin{equation*}
\begin{aligned}
\operatorname{Num}_4 &= -(\beta_1 + \beta_2) (2 x_4 + x_5) \operatorname{sin}(x_2)\\
& \quad - \beta_4 (2 x_4 + x_5 + x_6) \operatorname{sin}(x_2 + x_3) \\
\operatorname{Num}_5 &= -\beta_3 (2 x_4 + 2x_5 + x_6) \operatorname{sin}(x_3) \\
& \quad - \beta_4 (2 x_4 + x_5 + x_6) \operatorname{sin}(x_2 + x_3).
\end{aligned}
\end{equation*}
It can be easily shown that 
\begin{equation*}
\begin{aligned}
[g_1,g_2] &= 0 \\
[g_i,[f,g_j]] &= 0 \quad \operatorname{for} \, i,j=1,2,
\end{aligned}
\end{equation*}
and $[f,[f,g_1]]$, $[f,[f,g_2]]$ and $[[f,g_1],[f,g_2]]$ are linear combinations of $f$, $[f,g_1]$ and $[f,g_2]$. By using Lemma~\ref{lm:involutive}, the distribution $\varDelta$ is involutive, with the dimension of 5. We thus conclude that the three-link horizontal manipulator with configuration $3$ is not accessible from any state. 
\end{proof}

\begin{corollary}
The three-link horizontal manipulator with configuration 3 is not STLC from any state.
\end{corollary}

\subsection{N-link Analysis}
In this section, we will extend the previous conclusions for a three-link manipulator to a more general case, \textit{i.e.},  for an $N$-link ($N \geq 3$) manipulator. Note that the previous results can be used to verify the following computations.

\begin{theorem}
For an $N$-link ($N \geq 3$) horizontal manipulator with one unactuated joint, if the first joint is actuated, it  is accessible from almost any state, and also STLC from a subset of equilibrium points based on Sussmann's general theorem for STLC \cite{sussmann1987general}. Otherwise, it is neither accessible nor STLC from any state.
\end{theorem}

\begin{figure}
    \centering
    \includegraphics[width = .4\textwidth]{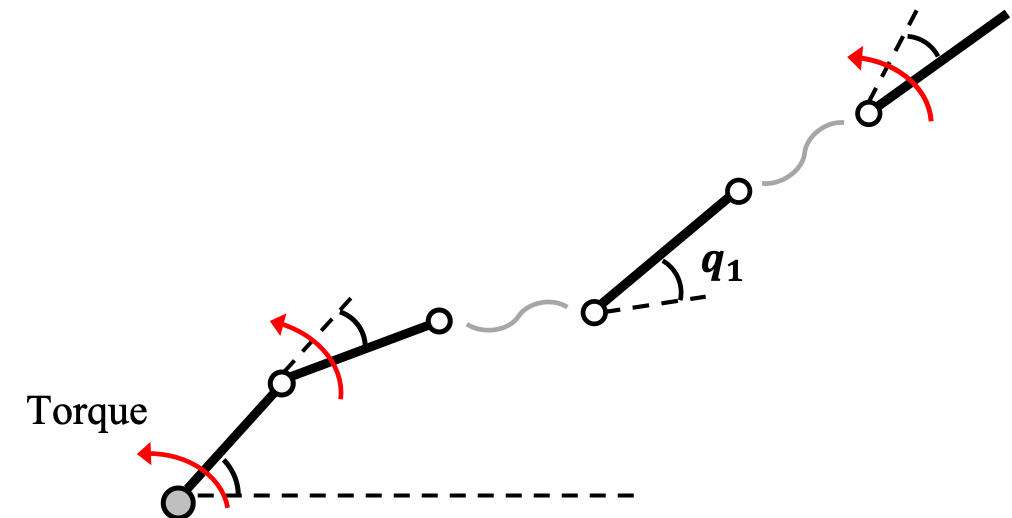}
    \caption{An $N$-Link horizontal manipulator with only one unactuated joint. $q_1$ is used to label the unactuated joint. If that joint is not the first (base) joint, the manipulator is accessible from almost any state, and also STLC from a subset of equilibrium points. Otherwise it is neither accessible nor STLC from any state.}
    \label{fig:nlink}
\end{figure}

\begin{proof}
The $N$-Link horizontal model is illustrated in Fig.~\ref{fig:nlink}. The unactuated joint can be any among the $N$ joints. For convenience of expressions, we always use $q_1$ to denote the unactuated joint. The other actuated joints are thus denoted as $q_2,\dotsc,q_n$, respectively. Define the configuration states $q \triangleq \left(q_1\,,q_2,\,\dotsc,\,q_n\right)$ and the velocity states $\dot{q}\triangleq (\dot{q}_1,\,\dot{q}_2,\,\dotsc,\,\dot{q}_n)$. For a general $N$-link model, the kinetic energy is
$T = (1/2) M_{ij}(q) \dot{q}_i \dot{q}_j$ where $i,j = 1,2,\dots,n$, so

\begin{equation}
\frac{d}{dt}\left(\frac{\partial{T}}{\partial{\dot{q}_k}}\right)-\frac{\partial{T}}{\partial{q_k}} = \tau_k \quad k = 1,2,\dotsc,n
\end{equation}

\begin{equation}
\label{eq:lag2}
\frac{d}{dt}\left(\frac{1}{2}M_{kj}\dot{q}_j+\frac{1}{2}M_{ik}\dot{q}_i\right)-\frac{1}{2}\frac{\partial{M_{ij}}}{\partial{q_k}}\dot{q}_i\dot{q}_j = \tau_k.
\end{equation}

Using symmetry of $M_{ij}$, simplifies Eq.~(\ref{eq:lag2})

\begin{equation}
\frac{d}{dt}\left(M_{ik}\dot{q}_i\right)-\frac{1}{2}\frac{\partial{M_{ij}}}{\partial{q_k}}\dot{q}_i\dot{q}_j = \tau_k
\end{equation}

\begin{equation}
M_{ik}\ddot{q}_i+\frac{\partial{M_{ik}}}{\partial{q_j}}\dot{q}_i\dot{q}_j-\frac{1}{2}\frac{\partial{M_{ij}}}{\partial{q_k}}\dot{q}_i\dot{q}_j = \tau_k.
\end{equation}

Using partial feedback linearization, we introduce new inputs $u_2,\dotsc,u_n$ and construct the control inputs by
\iffalse
\begin{equation}
\ddot{q}_1 = \frac{1}{M_{11}}\left(\frac{1}{2}\frac{\partial M_{ij}}{\partial q_1}\dot{q}_i\dot{q}_j-\frac{\partial M_{i1}}{\partial q_j}\dot{q}_i\dot{q}_j-\sum_{l=2}^{l=n}M_{i1}u_i\right)
\end{equation}
\fi

\begin{equation}
\begin{aligned}
\tau_k &= \frac{M_{1k}}{M_{11}}\left(\frac{1}{2}\frac{\partial M_{ij}}{\partial q_1}\dot{q}_i\dot{q}_j-\frac{\partial M_{i1}}{\partial q_j}\dot{q}_i\dot{q}_j-\sum_{l=2}^{l=n}M_{l1}u_l\right) \\
&\quad +\sum_{l=2}^{l=n}M_{lk}u_l+\frac{\partial{M_{ik}}}{\partial{q_j}}\dot{q}_i\dot{q}_j-\frac{1}{2}\frac{\partial{M_{ij}}}{\partial{q_k}}\dot{q}_i\dot{q}_j,
\end{aligned}
\end{equation}
which give
\begin{equation}
\begin{aligned}
\ddot{q}_1 &= \frac{1}{M_{11}}\left(\frac{1}{2}\frac{\partial M_{ij}}{\partial q_1}\dot{q}_i\dot{q}_j-\frac{\partial M_{i1}}{\partial q_j}\dot{q}_i\dot{q}_j-\sum_{l=2}^{l=n}M_{i1}u_i\right) \\
\ddot{q}_2 &= u_2 \\
& \quad \vdots \\
\ddot{q}_n &= u_n.
\end{aligned}
\end{equation}

Therefore, the equation of dynamics for a general $N$-Link model is
\begin{equation}
\dot{x} = f(x)+g_2(x)u_2+\dotsc+g_n(x)u_n,
\end{equation}
where the states
$x =
(q_1,\,q_2,\,\dots,\,q_n,\,\dot{q}_1,\,\dot{q}_2,\,\dots,\,\dot{q}_n)$, and 
\begin{equation}
\label{eq:vecfields}
f(x) = 
\mleft[
\begin{array}{c}
\dot{q}_1 \\
\dot{q}_2 \\
\vdots \\
\dot{q}_n \\
\hline
\hat{f}(q,\dot{q}) \\
0 \\
\vdots \\
0\\
0 \\
0 \\
\vdots \\
0
\end{array}
\mright]
\quad 
g_a(x) = 
\mleft[
\begin{array}{c}
0 \\
0 \\
\vdots \\
0 \\
\hline
\hat{g}_a(q) \\
0 \\
\vdots \\
0 \\
1 \\
0 \\
\vdots \\
0
\end{array}
\mright]
\begin{aligned}
& \quad a = 2,\,\dots,\, n \\[50pt]
&\leftarrow \text{$(n+a)$-th row}
\end{aligned}
\end{equation}

\begin{equation}
\label{eq:driftfield}
\hat{f}(q,\dot{q}) = \frac{1}{M_{11}}\left(\frac{1}{2}\frac{\partial M_{ij}}{\partial q_1}\dot{q}_i\dot{q}_j-\frac{\partial M_{i1}}{\partial q_j}\dot{q}_i\dot{q}_j\right)
\end{equation}

\begin{equation}
\label{eq:controlfield}
\hat{g}_a(q) = -\frac{M_{a1}}{M_{11}}.
\end{equation}

The Lie brackets are given by
\begin{equation}
\begin{aligned}
\label{eq:threevecfields}
[f,g_a](x) = 
\mleft[
\begin{array}{c}
-\hat{g}_a(q) \\
0 \\
\vdots \\
0 \\
-1 \\
0 \\
\vdots \\
0 \\
\hline
P_a(x) \\
0 \\
\vdots \\
0
\end{array}
\mright]
&
\begin{aligned}
\leftarrow{\text{$a$-th}}\\
\quad {\text{row}}
\\[30pt]
\end{aligned}
[g_a,[f,g_b]](x)=
\mleft[
\begin{array}{c}
0 \\
0 \\
\vdots \\
0 \\
0 \\
0 \\
\vdots \\
0 \\
\hline
P_{ab}(x) \\
0 \\
\vdots \\
0
\end{array}
\mright] \\
[f,[g_a,[f,g_b]]](x) &= 
\mleft[
\begin{array}{c}
-P_{ab}(x) \\
0 \\
\vdots \\
0\\
\hline
* \\
0 \\
\vdots \\
0
\end{array}
\mright],
\end{aligned}
\end{equation}
where $a,\,b = 2,\,\dots,\,n$, an arbitrary expression is represented by $*$ and
\begin{equation}
\label{eq:Pa}
\begin{aligned}
P_a(x) = \frac{\partial{\hat{g}_a(q)}}{\partial{q}_i} \dot{q}_i-\frac{\partial{\hat{f}(q,\dot{q})}}{\partial{\dot{q}_a}}-\frac{\partial{\hat{f}(q,\dot{q})}}{\partial{\dot{q}_1}}\hat{g}_a(q)
\end{aligned}
\end{equation}
\begin{equation}
\label{eq:Pab}
\begin{aligned}
P_{ab}(x) &= \underbrace{\frac{\partial{\hat{g}_b(q)}}{\partial{q_a}} + \frac{\partial{\hat{g}_a(q)}}{\partial{q_b}}}_{\circled{1}}\underbrace{-\frac{\partial^2\hat{f}(q,\dot{q})}{\partial{\dot{q}_a}\partial{\dot{q}_b}}}_{\circled{2}}\underbrace{-\hat{g}_a(q)\hat{g}_b(q)\frac{\partial^2 f(q,\dot{q})}{\partial \dot{q}_1^2}}_{\circled{3}} \\
 & \underbrace{\quad - \hat{g}_a(q)\frac{\partial^2 \hat{f}(q,\dot{q})}{\partial \dot{q}_1 \dot{q}_b}}_{\circled{4}}\underbrace{-\hat{g}_b(q)\frac{\partial^2 \hat{f}(q,\dot{q})}{\partial \dot{q}_1 \partial \dot{q}_a}}_{\circled{5}} \\
 & \quad \underbrace{+ \hat{g}_a(q)\frac{\partial \hat{g}_b(q)}{\partial q_1}}_{\circled{6}} \underbrace{+ \hat{g}_b(q) \frac{\partial \hat{g}_a(q)}{\partial q_1}}_{\circled{7}}.
\end{aligned}
\end{equation}

Substituting Eq. (\ref{eq:driftfield}) and (\ref{eq:controlfield}) into Eq. (\ref{eq:Pa}), the three terms in Eq.~(\ref{eq:Pa}) are computed in order,
\begin{equation*}
\begin{aligned}
\frac{\partial{\hat{g}_a(q)}}{\partial{q}_i} \dot{q}_i 
&= -\frac{1}{M_{11}^2}\left(M_{11}\frac{\partial M_{a1}}{\partial q_i}-M_{a1}\frac{\partial M_{11}}{\partial q_i}\right) \dot{q}_i \\
-\frac{\partial{\hat{f}(q,\dot{q})}}{\partial{\dot{q}_a}} 
&= \frac{1}{M_{11}}\frac{\partial M_{a1}}{\partial q_i}\dot{q}_i+\frac{1}{M_{11}}\frac{\partial M_{i1}}{\partial q_a}\dot{q}_i - \frac{1}{M_{11}}\frac{\partial M_{ia}}{\partial q_1}\dot{q}_i \\
-\frac{\partial{\hat{f}(q,\dot{q})}}{\partial{\dot{q}_1}}\hat{g}_a(q) 
&= \left(\frac{1}{M_{11}}\frac{\partial M_{11}}{\partial q_i}\dot{q}_i+\frac{1}{M_{11}}\frac{\partial M_{i1}}{\partial q_1}\dot{q}_i\right)\left(-\frac{M_{a1}}{M_{11}}\right) \\
& \quad - \frac{1}{M_{11}}\frac{\partial M_{i1}}{\partial q_1}\dot{q}_i\left(-\frac{M_{a1}}{M_{11}}\right) \\
& = -\frac{M_{a1}}{M_{11}^2}\frac{\partial M_{11}}{\partial q_i}\dot{q}_i.
\end{aligned}
\end{equation*}
Summing the above three terms leads to
\begin{equation}
\label{eq:Pa_simplify}
\begin{aligned}
P_a(x) &= \frac{1}{M_{11}}\frac{\partial M_{i1}}{\partial q_a}\dot{q}_i - \frac{1}{M_{11}}\frac{\partial M_{ia}}{\partial q_1}\dot{q}_i.
\end{aligned}
\end{equation}

Similarly, the seven terms in Eq. (\ref{eq:Pab}) are, \\
\begin{equation*}
\begin{aligned}
%\circled{1} \\
\frac{\partial{\hat{g}_b(q)}}{\partial{q_a}} + \frac{\partial{\hat{g}_a(q)}}{\partial{q_b}} 
&=  -\frac{1}{M_{11}^2}\left(M_{11}\frac{\partial M_{b1}}{\partial q_a}-M_{b1}\frac{\partial M_{11}}{\partial q_a}\right)  \\ & \quad -\frac{1}{M_{11}^2}\left(M_{11}\frac{\partial M_{a1}}{\partial q_b}-M_{a1}\frac{\partial M_{11}}{\partial q_b}\right),\\
%\circled{2}
-\frac{\partial^2\hat{f}(q,\dot{q})}{\partial{\dot{q}_a}\partial{\dot{q}_b}} 
&= \frac{1}{M_{11}}\frac{\partial}{\partial \dot{q}_a}\left(\frac{\partial M_{b1}}{\partial q_j}\dot{q}_j + \frac{\partial M_{i1}}{\partial q_b}q_i-\frac{\partial M_{ib}}{\partial q_1}\dot{q}_i\right) \\
& = \frac{1}{M_{11}}\left(\frac{\partial M_{b1}}{\partial q_a}+\frac{\partial M_{a1}}{\partial q_b}-\frac{\partial M_{ab}}{\partial q_1}\right),\\
%\circled{3}
-\hat{g}_a(q)\hat{g}_b(q)\frac{\partial^2 f(q,\dot{q})}{\partial \dot{q}_1^2} 
&= -\frac{M_{a1}}{M_{11}}\frac{M_{b1}}{M_{11}}\left(-\frac{1}{M_{11}}\frac{\partial M_{11}}{\partial q_1}\right) \\
& = \frac{M_{a1}M_{b1}}{M_{11}^3}\frac{\partial M_{11}}{\partial q_1},\\
%\circled{4}
- \hat{g}_a(q)\frac{\partial^2 \hat{f}(q,\dot{q})}{\partial \dot{q}_1 \dot{q}_b} 
&= -\frac{M_{a1}}{M_{11}}\left(\frac{1}{M_{11}}\frac{\partial M_{11}}{\partial q_b}\right) \\
& = -\frac{M_{a1}}{M_{11}^2}\frac{\partial M_{11}}{\partial q_b},\\
\end{aligned}
\end{equation*}
\begin{equation*}
\begin{aligned}
%\circled{5}
- \hat{g}_b(q)\frac{\partial^2 \hat{f}(q,\dot{q})}{\partial \dot{q}_1 \dot{q}_a} &= -\frac{M_{b1}}{M_{11}}\left(\frac{1}{M_{11}}\frac{\partial M_{11}}{\partial q_a}\right) \\
& = -\frac{M_{b1}}{M_{11}^2}\frac{\partial M_{11}}{\partial q_a},\\
%\circled{6}
\hat{g}_a(q)\frac{\partial \hat{g}_b(q)}{\partial q_1} &= -\frac{M_{a1}}{M_{11}}\left(-\frac{1}{M_{11}^2}\left(M_{11}\frac{\partial M_{b1}}{\partial q_1}-M_{b1}\frac{\partial M_{11}}{\partial q_1}\right)\right) \\
& = \frac{M_{a1}}{M_{11}^2}\frac{\partial M_{b1}}{\partial q_1}-\frac{M_{a1}M_{b1}}{M_{11}^3}\frac{\partial M_{11}}{\partial q_1},\\
%\circled{7}
\hat{g}_b(q)\frac{\partial \hat{g}_a(q)}{\partial q_1} &= -\frac{M_{b1}}{M_{11}}\left(-\frac{1}{M_{11}^2}\left(M_{11}\frac{\partial M_{a1}}{\partial q_1}-M_{a1}\frac{\partial M_{11}}{\partial q_1}\right)\right) \\
& = \frac{M_{b1}}{M_{11}^2}\frac{\partial M_{a1}}{\partial q_1}-\frac{M_{b1}M_{a1}}{M_{11}^3}\frac{\partial M_{11}}{\partial q_1}.
\end{aligned}
\end{equation*}
Summing the above seven terms yields to
\begin{equation}
\label{eq:Pab_simplify}
\begin{aligned}
P_{ab}(x) &= \frac{1}{M_{11}^2}\frac{\partial (M_{a1}M_{b1})}{\partial q_1} - \frac{1}{M_{11}}\frac{\partial M_{ab}}{\partial q_1} \\
& \quad -\frac{M_{a1}M_{b1}}{M_{11}^3}\frac{\partial M_{11}}{\partial q_1}.
\end{aligned}
\end{equation}

\textbf{Case 1: when the first joint is actuated}

Observe that the $P_{ab}(x)$ in Eq.~(\ref{eq:Pab_simplify}) is only dependent on the configuration states $q$. We pick $2n$ vector fields as follows,
\begin{equation}
\label{eq:2nvec}
\centering
\begin{aligned}
\underbrace{
\mleft[
\begin{array}{c}
0 \\
0 \\
\vdots \\
0 \\
\hline
P_{ab}(x) \\
0 \\
0 \\
0 \\
\vdots \\
0 \\
0
\end{array}
\mright]
}_{[g_a,[f,g_b]]}
\quad
\underbrace{
\mleft[
\begin{array}{c}
0 \\
0 \\
\vdots \\
0 \\
\hline
\hat{g}_2(q) \\
1 \\
0 \\
0 \\
\vdots \\
0 \\
0
\end{array}
\mright]
}_{g_2}
\quad
\underbrace{
\mleft[
\begin{array}{c}
0 \\
0 \\
\vdots \\
0 \\
\hline
\hat{g}_3(q) \\
0 \\
1 \\
0 \\
\vdots \\
0 \\
0
\end{array}
\mright]
}_{g_3}
\dotsc
\underbrace{
\mleft[
\begin{array}{c}
0 \\
0 \\
\vdots \\
0 \\
\hline
\hat{g}_n(q) \\
0 \\
0 \\
0 \\
\vdots \\
0 \\
1
\end{array}
\mright]
}_{g_n}\\
\underbrace{
\mleft[
\begin{array}{c}
-P_{ab}(x) \\
0 \\ 
0 \\
0 \\
\vdots \\
0\\
0\\
\hline
* \\
0 \\
\vdots \\
0 
\end{array}
\mright] 
}_{[f,[g_a,[f,g_b]]]}
\quad
\underbrace{
\mleft[
\begin{array}{c}
-\hat{g}_2(q) \\
-1 \\ 
0 \\
0 \\
\vdots \\
0\\
0\\
\hline
P_2(x) \\
0 \\
\vdots \\
0 
\end{array}
\mright]
}_{[f,g_2]}
\quad
\underbrace{
\mleft[
\begin{array}{c}
-\hat{g}_3(q) \\
0 \\ 
-1 \\
0 \\
\vdots \\
0\\
0\\
\hline
P_3(x) \\
0 \\
\vdots \\
0 
\end{array}
\mright]
}_{[f,g_3]}
\dotsc
\underbrace{
\mleft[
\begin{array}{c}
-\hat{g}_n(q) \\
0 \\ 
0 \\
0 \\
\vdots \\
0\\
-1\\
\hline
P_n(x) \\
0 \\
\vdots \\
0 
\end{array}
\mright]
}_{[f,g_n]},
\end{aligned}
\end{equation}
where $a$ and $b$ are picked randomly from a set $\{2,\,\dotsc,\,n\}$.

When the $P_{ab}$ is non-zero, it is easy to show that 
\begin{equation*}
\begin{aligned}
&\gamma_1 [g_a,[f,g_b]]+\gamma_2  g_2 +\gamma_3  g_3 + \dots + \gamma_n  g_n +\gamma_{n+1}  [f,[g_k,[f,g_l]]] \\
& \quad +\gamma_{n+2}  [f,g_2] +\gamma_{n+3}  [f,g_3] + \dots + \gamma_{2n} [f,g_n] = 0
\end{aligned}
\end{equation*}
is satisfied only when $\gamma_1 = \gamma_2 = \dotsc = \gamma_{2n} = 0$, which indicates that the $2n$ vector fields are linearly independent and span a $2n$-dimensional space. Thus, the $N$-link horizontal pendubot is accessible from almost any state. Next we consider STLC.

Since in the $2n$ vector fields in Eq.~(\ref{eq:2nvec}), $a$ and $b$ can be picked arbitrarily from the set $\{2,\,\dotsc,\,n\}$, we further require that $a \neq b$ to obtain $2n$ good brackets. Moreover, we need to verify whether all bad brackets can be $\theta$-neutralized by these good brackets. Following the same analysis for a three-link pendubot, we first specify a control vector field $g_a$, and assign 1 as the $\theta$-degree to the vector fields $f$ and $g_a$. All other control vector fields have 2 as the $\theta$-degree. Therefore, the bad brackets 
\begin{equation*}
[g_m,[f,g_m]] \,\operatorname{where}\, m = 2,\,...,\,n \; \operatorname{and}\; m\neq a,
\end{equation*}
can be neutralized by the good bracket $[g_a,[f,g_m]]$.

However, the bad bracket $[g_a,[f,g_a]]$ has lower degree than the good one. We have to further make $[g_a,[f,g_a]]$ trivial, which is to require the $P_{aa}(x)=0$ in $[g_a,[f,g_a]]$. Therefore, the maximum $\theta$-degree for the $2n$ good brackets is 5.

Using Theorem~\ref{nontrivialbrackets}, all other nontrivial bad brackets (evaluated at the equilibrium) have a $\theta$-degree larger than 5, and can thus be $\theta$-neutralized easily. Because the specified $g_a$ can be any vector field among $\{g_2,\,\dotsc,\,g_n\}$, it is concluded that the system is STLC from any equilibrium state $x_e$ satisfying 
\begin{equation*}
\begin{aligned}
& 1)\,P_{ab}(x_e) \neq 0 \\
& 2)\, P_{aa}(x_e)=0\, \operatorname{or} \, P_{bb}(x_e) = 0,
\end{aligned}
\end{equation*}
for some $a,\,b \in \{2,\,\dotsc,\,n\}$ and $a \neq b$.

\textbf{Case 2: when the first joint is unactuated} 

The unactuated joint $q_1$ is ankle joint, which does not appear in the inertia matrix. Thus we have the fact that
\begin{equation*}
\frac{\partial M_{ij}}{\partial q_1} = 0 \quad i,j=1,2\dotsc,n,
\end{equation*}
which leads to
\begin{equation}
P_a(x) = \frac{1}{M_{11}}\frac{\partial M_{i1}}{\partial q_a}\dot{q}_i 
\end{equation}
\begin{equation}
\label{eq:Pabacrobat}
P_{ab}(x) = 0.
\end{equation}

Thus we conclude that the vector fields $[g_a,[f,g_b]]$ in Eq. (\ref{eq:threevecfields}) are trivial. Furthermore, it can be easily computed for the model that
\begin{equation*}
[f,[f,g_a]](x) = 
\mleft[
\begin{array}{c}
* \\
0 \\
\vdots \\
0\\
\hline
Q_a(x) \\
0 \\
\vdots \\
0
\end{array}
\mright],
\quad
a = 2,\dotsc,n,
\end{equation*}
where 

\begin{equation}\
\label{eq:ffg_a}
\begin{aligned}
Q_a(x) & = \underbrace{\frac{\partial P_a(x)}{\partial q_i}\dot{q}_i}_{\circled{1}}\underbrace{+\frac{\partial P_a(x)}{\partial \dot{q}_i}\hat{f}(q,\dot{q})}_{\circled{2}} \\
& \underbrace{+\frac{\partial \hat{f}(q,\dot{q})}{q_1}\hat{g}_a(q)}_{\circled{3}}\underbrace{+\frac{\partial \hat{f}(q,\dot{q})}{\partial q_a}}_{\circled{4}}\underbrace{-\frac{\partial \hat{f}(q,\dot{q})}{\partial \dot{q}_1}P_a(x)}_{\circled{5}},\\
 & \quad i = 1,2,\dotsc,n.
\end{aligned}
\end{equation}

The five terms in Eq.~(\ref{eq:ffg_a}) are, \\
%\circled{1} \\
\begin{equation*}
\begin{aligned}
\frac{\partial P_a(x)}{\partial q_i}\dot{q}_i &= \left(\frac{1}{M_{11}}\frac{\partial ^2 M_{j1}}{\partial q_a \partial q_i}-\frac{1}{M_{11}^2}\frac{\partial M_{11}}{\partial q_i}\frac{\partial M_{j1}}{\partial q_a}\right)\dot{q}_i\dot{q}_j,\\
%\circled{2}
\frac{\partial P_a(x)}{\partial \dot{q}_1}\hat{f}(q,\dot{q}) &= -\frac{1}{M_{11}^2}\frac{\partial M_{11}}{\partial q_a}\frac{\partial M_{i1}}{\partial q_j}\dot{q}_i\dot{q}_j,\\
%\circled{3}
\frac{\partial \hat f(q,\dot{q})}{q_1}\hat{g}_a(q) &= 0,\\
%\circled{4}
\frac{\partial \hat{f}(q,\dot{q})}{\partial q_a} &= \left(\frac{1}{M_{11}^2}\frac{\partial M_{11}}{\partial q_a}\frac{\partial ^2 M_{i1}}{\partial q_j}-\frac{1}{M_{11}}\frac{\partial M_{i1}}{\partial q_a \partial q_j}\right)\dot{q}_i\dot{q}_j,\\
%\circled{5}
-\frac{\partial \hat{f}(q,\dot{q})}{\partial \dot{q}_1}P_a(x) &= \frac{1}{M_{11}^2}\frac{\partial M_{i1}}{\partial q_a}\frac{\partial M_{11}}{\partial q_j}\dot{q}_i \dot{q}_j.
\end{aligned}
\end{equation*}

Summing the above terms yields to $Q_a(x)=0$, and thus the vector field $[f,[f,g_a]]$ is in the following form,
\begin{equation*}
[f,[f,g_a]](x) = 
\mleft[
\begin{array}{c}
* \\
0 \\
\vdots \\
0\\
\hline
0 \\
0 \\
\vdots \\
0
\end{array}
\mright].
\end{equation*}

Recall from Eq. (\ref{eq:threevecfields}) that 
\begin{equation}
\label{eq:fgbfga}
[f,[g_b,[f,g_a]]](x) = 
\mleft[
\begin{array}{c}
-P_{ba}(x) \\
0 \\
\vdots \\
0\\
\hline
* \\
0 \\
\vdots \\
0
\end{array}
\mright],
\end{equation}
where the $*$ in Eq. (\ref{eq:fgbfga}) is
\begin{equation*}
* = \frac{\partial P_{ba}}{\partial q_i}\dot q_i-P_{ba}\frac{\partial \hat{f}}{\partial \dot{q}_1},
\end{equation*}
and it leads to $[f,[g_b,[f,g_a]]]=0$ due to $P_{ba}=0$ in Eq.~(\ref{eq:Pabacrobat}). By using the Jacobi identity, we further have
\begin{equation}
[[f,g_a],[f,g_b]]+[f,[g_b,[f,g_a]]]+[g_b,[[f,g_a],f]]=0,
\end{equation}
which yields to
\begin{equation}
[[f,g_a],[f,g_b]] = [g_b,[f,[f,g_a]]].
\end{equation}
Thus it can be easily checked that the vector field $[[f,g_a],[f,g_b]]$ is also in the form,
\begin{equation*}
[[f,g_a],[f,g_b]](x) = 
\mleft[
\begin{array}{c}
* \\
0 \\
\vdots \\
0\\
\hline
0 \\
0 \\
\vdots \\
0
\end{array}
\mright],
\quad
a,b = 2,\dotsc,n.
\end{equation*}

Consider a linear combination of $f,\,[f,g_2],\,\dotsc,\,[f,g_n]$,
\begin{equation*}
f(x)+\sum_{a=2}^{a=n}\dot{q}_a \cdot [f,g_a](x)=
\mleft[
\begin{array}{c}
* \\
0 \\
\vdots \\
0\\
\hline
Q_2(x) \\
0 \\
\vdots \\
0
\end{array}
\mright],
\end{equation*}
where 
\begin{equation}
\begin{aligned}
Q_2(x) &= \hat f(q,\dot q) + \sum_{a=2}^{a=n}\dot{q}_a P_a(x) \\
& = -\frac{1}{M_{11}}\frac{\partial M_{i1}}{\partial q_j}\dot{q}_i\dot{q}_j+ \sum_{a=2}^{a=n}\dot{q}_a\left(\frac{1}{M_{11}}\frac{\partial M_{i1}}{\partial q_a}\dot{q}_i\right)\\
& = -\frac{1}{M_{11}}\frac{\partial M_{i1}}{\partial q_1}\dot{q}_i\dot{q}_1 = 0.
\end{aligned}
\end{equation}

Therefore, the vector fields $[f,[f,g_a]]$ and $[[f,g_a],[f,g_b]]$ are both linear combinations of $f,\,[f,g_2],\,\dotsc,\,[f,g_n]$, and the accessibility distribution for the model
\begin{equation}
\varDelta(x) = \operatorname{span}\{f,\,g_a,\,[f,g_a]\},\quad a = 2,\dotsc,n
\end{equation}
is involutive with a dimension of $2n-1$. Thus it is not accessible from any state $x \in \mathcal{M}$. The result is consistant with the fact of momentum conservation for an $N$-link horizontal manipulator with the first joint unactuated. It can be naturally extended that such model is not STLC either.

\end{proof}

\section{Conclusions}
\label{sec:conclusions}
This paper presents the accessibility and small-time local controllability (STLC) results for $N$-link horizontal planar manipulators with one unactuated joint. Different actuator configurations are considered. It exploits the Lie brackets to show that the two-link pendubot is accessible from almost any state but does not satisfy Sussmann's general theorem for STLC. In contrast, a two-link acrobat is not accessible from any state, which is due to the angular momentum conservation of the model. Furthermore, Lie brackets can also show that the acrobat starting from zero-velocity states has a codimension of two, which is due to that the second-order nonholonomic constraint is reduced to a holonomic constraint in this case.

As for $N$-link ($N \geq 3$) manipulators with one unactuated joint, it is found that the manipulator with the first joint actuated is accessible from almost any state and STLC from a subset of equilibrium points. Otherwise, the model is neither accessible nor STLC from any state.

Another important contribution of the paper is that it studies realistic $N$-link models and incorporates the dynamics to the controllability analysis by using partial feedback linearization and index notation, which turn out to give relatively simple forms for some nontrivial vector fields generated from Lie brackets. These expressions enable us to determine at which configurations the model may lose full rank condition for accessibility.

\subsection *{Acknowledgments}
\label{acknowledgement}

The partial support of the US National Science Foundation under grant
IIS-1527393 is gratefully acknowledged.
\section{appendix}
\label{sec:appendix}
\subsection{Three-link model details}
\label{sec:appendix-A}
To facilitate the model expressions for three-link horizontal manipulators, let~\cite{mahindrakar2005controllability},
\begin{equation*}
\begin{aligned}
\alpha_1 &= m_1 l_{1c}^2 + m_2 l_1^2 + m_3 l_1^2+I_1 \\
\alpha_2 &= m_2 l_{2c}^2+m_3 l_2^2 + I_2 \\
\alpha_3 &= m_3 l_{3c}^2+I_3 \\
\beta_1 &= m_2 l_1 l_{2c} \\
\beta_2 &= m_3 l_1 l_2 \\
\beta_3 &= m_3 l_2 l_{3c} \\
\beta_4 &= m_3 l_1 l_{3c}.
\end{aligned}
\end{equation*}

\subsubsection{Configuration 1} 
The dynamics are described by
\begin{equation}
\label{eq:threelinkconfig1}
\begin{bmatrix}
M_{11} & M_{12} & M_{13} \\
M_{21} & M_{22} & M_{23} \\
M_{31} & M_{32} & M_{33}
\end{bmatrix}
\begin{bmatrix}
\ddot{q}_1 \\
\ddot{q}_2 \\
\ddot{q}_3
\end{bmatrix}
+
\begin{bmatrix}
C_1 \\
C_2 \\
C_3
\end{bmatrix}
=
\begin{bmatrix}
\tau_1 \\
\tau_2 \\
0
\end{bmatrix},
\end{equation}
where \footnote{$s_2,\,s_3,\,s_{23},\,c_2,\,c_3,\,c_{23}$ are abbreviations for $\operatorname{sin}(q_2),\,\operatorname{sin}(q_3),\,\operatorname{sin}(q_2+q_3),\,\operatorname{cos}(q_2),\,\operatorname{cos}(q_3),\,\operatorname{cos}(q_2+q_3)$, respectively.}
\begin{equation*}
\begin{aligned}
M_{11} &= \alpha_1+\alpha_2+\alpha_3+2\beta_1c_2+2\beta_2c_2+2\beta_3c_3+2\beta_4c_{23} \\
M_{12} &= \alpha_2+\alpha_3+(\beta_1+\beta_2)c_2+2\beta_3c_3+\beta_4c_{23} \\
M_{13} &= \alpha_3+\beta_3c_3+\beta_4c_{23} \\
M_{21} &= M_{12} \\
M_{22} &= \alpha_2+\alpha_3+2\beta_3c_3\\
M_{23} &= \alpha_3+\beta_3c_3 \\
M_{31} &= M_{13} \\
M_{32} &= M_{23} \\
M_{33} &= \alpha_3 \\
C_1 &=  -(2 \dot{q}_1 + \dot{q}_2) \dot{q}_2 (\beta_1 + \beta_2)s_2 - \beta_3 (2 \dot{q}_1 + 2 \dot{q}_2 + \dot{q}_3) \dot{q}_3 s_3 \\
 & \quad -\beta_4 (\dot{q}_2 + \dot{q}_3) (2 \dot{q}_1 + \dot{q}_2 + \dot{q}_3) s_{23} \\
C_2 &= \dot{q}_1^2((\beta_1+\beta_2)s_2+\beta_4s_{23})-2\beta_3\dot{q}_1\dot{q}_3s_3-2\beta_3\dot{q}_2\dot{q}_3s_3 \\
 & \quad -\beta_3\dot{q}_3^2s_3 \\
C_3 &= \dot{q}_1^2(\beta_3s_3+\beta_4s_{23})+2\dot{q}_1\dot{q}_2\beta_3s_3+\beta_3\dot{q}_2^2s_3.
\end{aligned}
\end{equation*}

Considering partial feedback linearization, we introduce inputs $u_1$ and $u_2$, and design the control inputs $\tau_1$ and $\tau_2$ by
\begin{equation*}
\begin{aligned}
\tau_1 &= \frac{M_{11}M_{33}-M_{13}M_{31}}{M_{33}}u_1+\frac{M_{12}M_{33}-M_{13}M_{32}}{M_{33}}u_2 \\
 & \quad +\frac{M_{33}C_1-M_{13}C_3}{M_{33}} \\
\tau_2 &= \frac{M_{21}M_{33}-M_{23}M_{31}}{M_{33}}u_1+\frac{M_{22}M_{33}-M_{23}M_{32}}{M_{33}}u_2 \\
& \quad +\frac{M_{33}C_2-M_{23}C_3}{M_{33}}.
\end{aligned}
\end{equation*}

Thus, the dynamics equation~(\ref{eq:threelinkconfig1}) can be simplified as 
\begin{equation*}
\begin{bmatrix}
\ddot{q}_1 \\ \ddot{q}_2 \\ \ddot{q}_3
\end{bmatrix}
=
\begin{bmatrix}
u_1 \\ u_2 \\ -M_{31}/M_{33}u_1-M_{32}/M_{33}u_2-C_3/M_{33}
\end{bmatrix}.
\end{equation*}

\subsubsection{Configuration 2}
The dynamics are described by
\begin{equation}
\label{eq:threelinkconfig2}
\begin{bmatrix}
M_{11} & M_{12} & M_{13} \\
M_{21} & M_{22} & M_{23} \\
M_{31} & M_{32} & M_{33}
\end{bmatrix}
\begin{bmatrix}
\ddot{q}_1 \\
\ddot{q}_2 \\
\ddot{q}_3
\end{bmatrix}
+
\begin{bmatrix}
C_1 \\
C_2 \\
C_3
\end{bmatrix}
=
\begin{bmatrix}
\tau_1 \\
0 \\
\tau_2 
\end{bmatrix},
\end{equation}
where  the  inertia  and  Coriolis  matrices  are  the same  with those in Eq.~(\ref{eq:threelinkconfig1}). Introduce inputs $u_1$ and $u_2$, and design the control inputs $\tau_1$ and $\tau_2$ by
\begin{equation*}
\begin{aligned}
\tau_1 &= \frac{M_{11}M_{22}-M_{12}M_{21}}{M_{22}}u_1+\frac{M_{13}M_{22}-M_{12}M_{23}}{M_{22}}u_2 \\
 & \quad +\frac{M_{22}C_1-M_{12}C_2}{M_{22}} \\
\tau_2 &= \frac{M_{22}M_{31}-M_{21}M_{32}}{M_{22}}u_1+\frac{M_{22}M_{33}-M_{23}M_{32}}{M_{22}}u_2 \\
& \quad +\frac{M_{22}C_3-M_{32}C_2}{M_{22}},
\end{aligned}
\end{equation*}
which yields to
\begin{equation*}
\begin{bmatrix}
\ddot{q}_1 \\ \ddot{q}_2 \\ \ddot{q}_3
\end{bmatrix}
=
\begin{bmatrix}
u_1 \\ -M_{21}/M_{22}u_1-M_{23}/M_{22}u_2-C_2/M_{22} \\ u_2
\end{bmatrix}.
\end{equation*}

\subsubsection{Configuration 3}
The dynamics are described by
\begin{equation}
\label{eq:threelinkconfig3}
\begin{bmatrix}
M_{11} & M_{12} & M_{13} \\
M_{21} & M_{22} & M_{23} \\
M_{31} & M_{32} & M_{33}
\end{bmatrix}
\begin{bmatrix}
\ddot{q}_1 \\
\ddot{q}_2 \\
\ddot{q}_3
\end{bmatrix}
+
\begin{bmatrix}
C_1 \\
C_2 \\
C_3
\end{bmatrix}
=
\begin{bmatrix}
0 \\
\tau_1 \\
\tau_2 
\end{bmatrix},
\end{equation}
where  the  inertia  and  Coriolis  matrices  are  the same  with those in Eq.~(\ref{eq:threelinkconfig1}). Introduce inputs $u_1$ and $u_2$, and design the control inputs $\tau_1$ and $\tau_2$ by
\begin{equation*}
\begin{aligned}
\tau_1 &= \frac{M_{11}M_{22}-M_{12}M_{21}}{M_{11}}u_1+\frac{M_{11}M_{23}-M_{13}M_{21}}{M_{11}}u_2 \\
 & \quad +\frac{M_{11}C_2-M_{21}C_1}{M_{11}} \\
\tau_2 &= \frac{M_{11}M_{32}-M_{12}M_{31}}{M_{11}}u_1+\frac{M_{11}M_{33}-M_{13}M_{31}}{M_{11}}u_2 \\
& \quad +\frac{M_{11}C_3-M_{31}C_1}{M_{11}},
\end{aligned}
\end{equation*}
which yields to
\begin{equation*}
\begin{bmatrix}
\ddot{q}_1 \\ \ddot{q}_2 \\ \ddot{q}_3
\end{bmatrix}
=
\begin{bmatrix}
 -M_{12}/M_{11}u_1-M_{13}/M_{11}u_2-C_1/M_{11}\\ u_1 \\ u_2
\end{bmatrix}.
\end{equation*}

\subsection{Proof of Theorem~\ref{nontrivialbrackets}}
\label{sec:appendix-B}
\begin{prop}
\label{prop1}
Every element in the Lie algebra $\mathcal{C}$ is a linear combination of vector fields of the form
\begin{equation}
\label{eq:bracketform}
    [X_k,[X_{k-1},[\cdots,[X_2,X_1]\cdots]]],
\end{equation}
where $X_i \in \mathcal{C}$, $i = 1,\cdots,k$.
\end{prop}

It can be easily proven by induction. The proof can also be found in \cite{lewis1995aspects}.

Now consider a general $N$-link model as described in section~\ref{sec:threelink}. Let $\bold{X} = \{f,g_2,\dotsc,g_n\}$, and $\mathcal{C}$ is the Lie algebra generated by the set $\bold{X}$. Using Proposition \ref{prop1}, we only need to consider brackets of the form (\ref{eq:bracketform}). Define 
\begin{equation*}
\begin{aligned}
    Br^k(\bold{X}) &= \Big\{B \in Br(\bold{X}) \mid \sum_{i=1}^{n-1} \delta^i(B)+\delta^0(B) = k \Big\}, \\
    Br_l(\bold{X}) &= \Big\{B \in Br(\bold{X}) \mid \sum_{i=1}^{n-1} \delta^i(B)-\delta^0(B) = l \Big\}, \\
\end{aligned}
\end{equation*}
where $k$ is the common definition of the degree of $B$.

For a vector field containing polynomials, if the nonzero components all have the same polynomial degree, we define such degree as the \emph{polynomial degree of the vector field}. If all the components are zero, we define the polynomial degree of the vector field as $-1$. To illustrate the idea, consider the vector fields as follows,
\begin{equation*}
    V_1 = 
    \begin{bmatrix}
    0 \\ 0 \\ x^2
    \end{bmatrix},
    \quad
    V_2 = 
    \begin{bmatrix}
    0 \\ 0 \\ 0
    \end{bmatrix}
    \quad
    \text{and}
    \quad
    V_3 = 
    \begin{bmatrix}
    x \\ 0 \\ x^2
    \end{bmatrix},
\end{equation*}
where $V_1$ has a polynomial (about $x$) degree of $2$, $V_2$ has a polynomial degree of $-1$, and the polynomial degree of $V_3$ cannot be defined since it contains $x$ and $x^2$ that have different polynomial degrees.

\begin{lemma}
\label{lm:polydegree}
Consider all vector fields generated by taking the Lie brackets on $\bold{X} = \{f,g_2,\dotsc,g_n\}$. The velocity coordinates only appear as homogeneous polynomials, perhaps with coefficients that are a function of the configuration variables\footnote{For example, $\dot{q}_1^2\operatorname{sin}q_1+\dot{q}_2^2+\dot{q}_1\dot{q}_3q_6$ is a second-order homogeneous polynomial about the velocity coordinates.}. Split the vector fields into top half (horizontal component) and bottom half (vertical component). The horizontal and vertical components may have different polynomial (about the velocity coordinates) degrees, and should satisfy one of the following rules:

(i) Let $l \geq -1$, if the polynomial degree of the horizontal component is $l$, then the polynomial degree of the vertical component is $l+1$. Moreover, bracketing by $f$ increases the polynomial degree of both the horizontal and vertical components by one, and bracketing by $g_a,\,a = 2,\dotsc,n$ reduces the polynomial degree of the components by one.

(ii) If the polynomial degree of the horizontal component is $-1$, the polynomial degree of the vertical component is $-1$. In this case, the vector field is trivial.
\end{lemma}

\begin{remark}
As a quick check, the lemma \ref{lm:polydegree} is true for the vector fields in Eq.~(\ref{eq:vecfields}) and (\ref{eq:threevecfields}). The horizontal component of the vector field $f$ has a polynomial degree of $1$, and the vertical component of $f$ has a polynomial degree of $2$, which satisfy the rule (i). The horizontal component of the vector field $g_a,\,a = 2,\dotsc,n$ has a polynomial degree of $-1$, and the vertical component of $g_a$ has a polynomial degree of $0$, which also satisfy the rule (i). 

Moreover, when bracketing the vector field $g_a$ by $f$, we generate a new vector field $[f,g_a]$, of which the horizontal component has a polynomial degree of $0$ and the vertical component has a polynomial degree of $1$. Both the horizontal and vertical components of $[f,g_a]$ have one larger degree than those of $g_a$.

Furthermore, when bracketing the vector field $[f,g_b]$ by $g_a$, we generate a new vector field $[g_a,[f,g_b]]$, of which the horizontal component has a polynomial degree of $-1$ and the vertical component has a polynomial degree of $0$. Both the horizontal and vertical components of $[g_a,[f,g_b]]$ have one smaller degree than those of $[f,g_b]$.
\end{remark}

\begin{proof}
Define $v^i = \dot{q}^i$ and denote the bracket
\begin{equation*}
B = B^i_h(q,v)\frac{\partial}{\partial q^i} + B^i_v(q,v)\frac{\partial}{\partial v^i}
\quad 
\text{or}
\quad
    B = 
    \begin{bmatrix}
    B_h^i \\
    B_v^i
    \end{bmatrix}
\end{equation*}
where $B_h^i$ is the horizontal component applied on the configuration states, and $B_v^i$ is the vertical component applied on the velocity states.

Now suppose the vector field $B$ satisfies the rule (i). Define $B[l]$ as the horizontal component with polynomial (about $v^i$) degree of $l$ and $B[l+1]$ as the vertical component with polynomial (about $v^i$) degree of $l+1$. Thus, it can be written
\begin{equation*}
    B = 
    \begin{bmatrix}
    B[l] \\ B[l+1]
    \end{bmatrix} \quad
    f = 
    \begin{bmatrix}
    f[1] \\ f[2]
    \end{bmatrix} \quad
    g_a = 
    \begin{bmatrix}
    g_a[-1] \\ g_a[0]
    \end{bmatrix}.
\end{equation*}

Bracketing $B$ by $f$, 
\begin{equation*}
\begin{aligned}
[f,B] 
&=
\begin{bmatrix}
\frac{\partial B[l]}{\partial q} &  \frac{\partial B[l]}{\partial v}\\
\frac{\partial B[l+1]}{\partial q} &  \frac{\partial B[l+1]}{\partial v}
\end{bmatrix}
\begin{pmatrix}
f[1] \\ f[2]
\end{pmatrix}
-\begin{bmatrix}
0 & I \\
\frac{\partial f[2]}{\partial q} &  \frac{\partial f[2]}{\partial v}
\end{bmatrix}
\begin{pmatrix}
B[l] \\ B[l+1]
\end{pmatrix}.
\end{aligned}
\end{equation*}

In $[f,B]$, the horizontal component is 
\begin{equation*}
    \frac{\partial B[l]}{\partial q}f[1] + \frac{\partial B[l]}{\partial v} f[2]-B[l+1].
\end{equation*} 
$\partial B[l]/\partial q$ has a polynomial degree of $l$, and multiplying it by $f[1]$ increases its degree to $l+1$. Similarly, the other two terms also have a polynomial degree of $l+1$. Thus, the horizontal component of $[f,B]$ has a polynomial degree of $l+1$. And the vertical component is
\begin{equation*}
    \frac{\partial B[l+1]}{\partial q}f[1] + \frac{\partial B[l+1]}{\partial v} f[2]-\frac{\partial f[2]}{\partial q}B[l]-\frac{\partial f[2]}{\partial v}B[l+1].
\end{equation*} 
 which has a polynomial degree of $l+2$. Thus, both the horizontal and vertical components of $[f,B]$ have one larger polynomial degree than those of $B$. 

Bracketing $B$ by $g_a$, 
\begin{equation*}
\begin{aligned}
[g_a,B] 
=
\begin{bmatrix}
\frac{\partial B[l]}{\partial q} &  \frac{\partial B[l]}{\partial v}\\
\frac{\partial B[l+1]}{\partial q} &  \frac{\partial B[l+1]}{\partial v}
\end{bmatrix}
\begin{pmatrix}
0 \\ g_a[0]
\end{pmatrix}
-\begin{bmatrix}
0 & 0 \\
\frac{\partial g_a[0]}{\partial q} &  0
\end{bmatrix}
\begin{pmatrix}
B[l] \\ B[l+1]
\end{pmatrix}.
\end{aligned}
\end{equation*}

In $[g_a,B]$, the horizontal component is
\begin{equation*}
    \frac{\partial B[l]}{\partial v}g_a[0],
\end{equation*}
which has a polynomial degree of $l-1$. And the vertical component is
\begin{equation*}
    \frac{\partial B[l+1]}{\partial v}g_a[0]-\frac{\partial g_a[0]}{\partial q}B[l],
\end{equation*}
which has a polynomial degree of $l$. Thus, both the horizontal and vertical components of $[g_a,B]$ have one smaller polynomial degree than those of $B$. 

Note that when $l = -1$ for $B$, simple computation may show that bracketing by $g_a$ will generate a new vector field that satisfies the rule (ii).

When the polynomial degrees of both the horizontal and vertical components are $-1$, which satisfies the rule (ii), the vector field is trivial in this case, and bracketing by either $f$ or $g_a$ returns a zero vector field, which still satisfies the rule (ii).

\end{proof}

\begin{lemma}
\label{lm:leq-1}
Let $k \geq 1$ be an integer, the bracket $B \in Br^k(\bold{X}) \cap Br_l(\bold{X})$ is zero at the equilibrium states for $l \leq -1$.
\end{lemma}

\begin{proof}
The lemma is true for $k=1$, in which $f$ is the only vector field satisfying $l \leq -1$. Suppose that the bracket $B$ is constructed by $m$ vector field $f$ and $m+l$ vector field $g_a$, $a = 2,\dotsc,n$. Using lemma \ref{lm:polydegree}, through bracketing by $f$ and $g_a$, the vector field $B$ can have the polynomial degree of $-l$ for the horizontal component and the polynomial degree of $1-l$ for the vertical component, or the vector field $B$ is a zero vector field. Since $l \leq -1$, for both the two cases, the bracket $B \in Br^k(\bold{X}) \cap Br_l(\bold{X})$ is zero at the equilibrium states for $l \leq -1$.

\end{proof}

\iffalse
\begin{lemma}
\label{lemma:eq1}
Let $k \geq 2$ be an integer, the bracket $B \in Br^k(\bold{X}) \cap Br_l(\bold{X})$ for $l = 1$ is always in the the form
\begin{equation}
\mleft[
\begin{array}{c}
0 \\
0 \\
\vdots \\
0 \\
\hline
p(q) \\
0 \\
\vdots \\
0
\end{array}
\mright],
\end{equation}
where $p(q)$ represents a polynomial of only the configuration states $q$. 
\end{lemma}

\begin{proof}
The result makes no sense when $k$ is even number, and is true for $k=3$, \textit{i.e.}, when $B = [g_a,[f,g_b]]$ (it can be checked that $[f,[g_a,g_b]]=0$). 

Now suppose the lemma is true for $k=3,\dotsc,j$ and let $B \in Br^{j+2}(\bold{X}) \cap Br_{1}(\bold{X})$ be of the form (\ref{eq:bracketform}). Then either $B=[f,[g_a,B']]$ or $B=[g_a,[f,B']]$, with $B' \in Br^j(\bold{X}) \cap Br_{1}(\bold{X})$ and $a = 2,\,\dots,\, n$.
\end{proof}
\fi

\begin{lemma}
\label{lm:geq2}
Let $k \geq 2$ be an integer, the bracket $B \in Br^k(\bold{X}) \cap Br_l(\bold{X})$ is identically zero for $l \geq 2$.
\end{lemma}

\begin{proof}
The proof follows the same line with the proof for lemma \ref{lm:leq-1}. Since there are at least two more vector fields $g_a$ than $f$ in constructing the bracket $B$, by using lemma \ref{lm:polydegree}, the generated bracket $B$ is identically zero.

\end{proof}

\addtolength{\textheight}{-0cm}

\bibliography{bibfile}
\bibliographystyle{IEEEtran}
\end{document}